# AXIOMATIC FOUNDATIONS OF PHYSICAL THEORIES

Stefano Bellucci[1,*] and Fabio Cardone[2]

1    INFN-Laboratori Nazionali di Frascati, Via E. Fermi 54, 00044 Frascati, Italy

2    National Research Council, Rome, Italy

*    Correspondence: stefano.bellucci@lnf.infn.it

ABSTRACT

In this paper we give a contribution to the taxonomy of physical theories. We provide here a thorough description of the axiomatic foundations of the most relevant physical theories, Mechanics, Special Relativity, General Relativity, Quantum Mechanics. The corresponding interactions will be dealt with as well, i.e. Gravity in the Minkowskian limit, Electricity without quantized energy, Gravity without quantized energy, Electricity with quantized energy. We pose the problem if the extension of the principle of solidarity to all interactions can impose to consider all variables as dynamic.

INTRODUCTION

All axiomatic foundations of physical theories have an inductive genesis and a deductive value *a posteriori*, if only for historical evidence.

Experimentation provides the information to induction; this is experimental physics.

Induction provides the elements for deduction; this is phenomenology in which known phenomena are put in order.

Deduction provides predictions amenable to experimentation; this is theoretical physics.

The physical theories from which we obtain predictions until now verified are four and are listed in correspondence to the fundamental interactions to which they can be applied.

| PHYSICAL THEORY | INTERACTION |
|---|---|
| Mechanics | Gravity in the Minkowskian limit |
| Special Relativity | Electricity without quantized energy |
| General Relativity | Gravity without quantized energy |
| Quantum Mechanics | Electricity with quantized energy |

DEFINITIONS

We now proceed to give definitions of the concepts that will be used limited to their meaning in physics.

Axiom: any assumption on which a mathematical theory such as geometry or real numbers is based, is otherwise known as a postulate.

Therefore axiom and postulate will be used synonymously.

Theoretical Physics: the description of natural phenomena in mathematical form.

Theory: in physics, an attempt to explain a certain class of phenomena as necessary consequences of other phenomena considered more primitive and less in need of explanation.

Principle: a scientific law that is highly general or fundamental and from which other laws are derived.

It is doubtful whether principles in physics can be considered as postulates, so the question arises whether to include a physical principle among the postulates of a theory.

METHODOLOGICAL AXIOMS

1 - Everything that is rational is not always real.

2 - The order and connection of ideas are identical to the order and connection of things (ordo et connexio idearum idem est ac ordo et connexio rerum), Baruch Spinoza (Ethica, II, pr. VII).

3 - The logic of nature is not human logic.

FIRST THEORY: MECHANICS

Mechanics can be founded on an axiomatic basis [1,2] using the principle of minimum action of Maupertuis [3,4], the conservation of total energy according to Hamilton [5] and the principle of relativity of Galilei [6,7], or, as can be demonstrated, it is the theory of gravitational interaction in the Minkowskian limit or flat space.

However, we prefer to illustrate an axiomatic foundation of Mechanics based on three principles which in a *a posteriori* view apply to bodies with constant mass different from zero, in this sense we can say that we speak of Corpuscular Mechanics.

In the statement is intended for natural motion that subject to conservative forces, or gravity and electricity, for adiabatic physical systems that do not exchange energy with the observer.

1 - Hamilton's principle or total energy principle

In natural motion with equal points of departure and arrival in space considered at synchronous times, that is equal, the total energy is conserved.

Here, we mean the points in the Euclidean space of Newtonian mechanics; this classical definition refers to nondissipative conservative interactions.

Note that the statement "at equal times" is well founded because in the Galilei-Newton transformations [8] for the temporal coordinate we have t = t' , that is the simultaneity in physical phenomena is allowed and valid.

2 - Maupertuis-Hoelder principle or stationary action principle

In isoenergetic natural motions, with equal instants of departure and arrival in time, considered at asynchronous times, i.e. different, the product of energy for time, called action, is conserved.

3 - Principle of minimum action or Euler's principle [9]

If the natural motion is adiabatic, the product of energy for time, the action, has always the minimum possible value.

It is appropriate here to introduce a digression. On the third principle we can graft the quantum hypothesis of Planck [10] and Sommerfeld [11,12] and indicate that this minimum value is always an integer multiple of a constant quantity $h$ for any physical system.

Always using a *a posteriori* vision, not historical, for bodies assumed massless at rest we can well extend the axiomatic foundation of Mechanics to the principle of minimum time of Fermat: the natural motion of light between two points in space is such that the corresponding time interval has always the minimum possible value.

This principle is at the base of the identification of the Geometric Optics with the Corpuscular Mechanics whose foundations have been stated before.

SECOND THEORY: RESTRICTED RELATIVITY

The Special Relativity has historically supplanted the Mechanics, incorporating it as its limit, as will be explained later.

In addition to its axiomatic foundation based on the two phenomenological postulates, it is possible to give an axiomatic statement, not phenomenological, of Special Relativity based on the equivalence of inertial systems, historically defined as a principle, and covariance (i.e. the laws of physics have the same mathematical form in all inertial systems).

We present here the three axioms structure according to Mignani – Recami [13].

### 1st Axiom - Properties of space - time

Space and time are homogeneous and space is isotropic.

### 2nd Axiom - Principle of Relativity

All physical laws are covariant when you go to observe them from an inertial system to a system moving at constant speed with respect to it.

### 3rd Axiom - Principle of Reinterpretation

Causality is realized by physical signals that are actually carried by objects having only finite positive energy.

This formulation exalts even more the contradictions between the Special Relativity and the non relativistic Quantum Mechanics which, because of its fundamental Schroedinger equation [14,15], violates quietly the 1st and 2nd Axiom. This because of the mathematical structure of this equation that treats in different way the variations of space and time through the derivation operations. As if it was not enough, the Relativistic Quantum Mechanics would violate the 3rd, with the negative energy, if there were not the experimental evidence of antimatter according to Dirac [16,17] and the renormalization technique according to Feynman [18,19]. The equivalence of the Feynman approach with that of Schwinger was shown in [20].

For the extension of the latter method to the nonabelian gauge fields [21].

The fundamental consequence of this three axioms formulation is that there is an invariant u having the physical dimensions of a velocity squared and finite value (although unlimited), which must be identified experimentally. From this last fact descends the principle of correspondence (of Special Relativity) according to which, if the invariant assumes an infinite value, we obtain the Mechanics as a consequence of Special Relativity.

It is interesting to note here that also the non-relativistic Quantum Mechanics has its own principle of correspondence with Mechanics.

The consequence of the existence of an invariant is the introduction in the physical theories of the concept of invariance or symmetry, in fact every invariant can be matched with a symmetry which becomes synonymous of invariance.

We indicate with symmetry the property of a quantity or physical law to remain unchanged as a result of certain operations or transformations.

Examples are the inertial mass and the elementary electric charge. A further example, which goes under the name of the CPT theorem, is the simultaneous combination of electric charge conjugation (i.e. reversal of the sign of charge), parity conjugation (i.e. spatial inversion or reflection) and time conjugation (i.e. time inversion or reflection, understood in mathematical sense) [22, 23].

While we indicate with invariance principle or symmetry law any principle for which a physical quantity or law has invariance under certain transformations, so it is common to use the name of such transformations to indicate the corresponding symmetry. Examples are Galilei transformations, symmetry by translation, and Lorentz transformations, symmetry by roto-translation, see [24,25] for a historical development, see [26] for a complete modern description of this topic.

Not always a symmetry can correspond to real physical objects, an example being the so called quarks.

THIRD THEORY: GENERAL RELATIVITY

Also the General Relativity [27,28] has its agile axiomatic foundation, with three axioms, but of different nature, from the genesis of the triads of those of Mechanics and Special Relativity, that initially is of phenomenological nature and then more properly of logical nature.

We could say, with Einstein, that the axioms of General Relativity, more than satisfying a phenomenological necessity, are meeting an aesthetic taste of logic.

### 1st Axiom - Principle of Equivalence

The position-dependent effects of gravity, i.e. local ones, are indistinguishable from those of a non-inertial reference system.

### 2nd Axiom - Covariance Principle

The laws of physics have the same mathematical form in all conceivable curvilinear coordinate systems.

### 3rd Axiom - Principle of Invariance

The laws of motion are the same in all reference systems, whether they are of variable speed or non-variable speed.

Notice that the laws of physics, the interactions, have as their consequence the laws of motion, the geodesics.

It is easy to see in the 3rd Axiom the crucial role of invariance, introduced as a consequence in the Special Relativity, as well as the logical - aesthetic taste of the 1st and 2nd Axiom, especially in the 1st one, which raises a fact considered accidental in Mechanics to the level of fundamental Axiom; on the other hand, the 3rd Axiom continues to yield a guarantee of Mechanics also in this new field of General Relativity.

The fundamental consequence of this axiomatic foundation is the "principle of solidarity", which is illustrated here according to Geymonat, and the consequent Cartesian negation of the physical vacuum by Einstein (in order to have this fundamental consequence we have avoided to illustrate this principle in the formulation according to Finzi). The space-time does not possess everywhere the same geometric structure independent from the field of interactions existing there (locally); it is the masses and the energy to curl (deform) variously the space-time, establishing a "solidarity" between the phenomena and the same space-time in which the phenomena take place.

Instead Mechanics, Special Relativity and Quantum Mechanics attribute an autonomous existence to space-time.

This latter idea, according to Einstein, can be "drastically expressed in this way : if matter should disappear, space and time would still remain, as a kind of stage (not deformable) for physical events".

General Relativity, on the other hand, argues that space-time does not have a separate existence from what fills it, i.e. there is no space-time without an interaction field.

Einstein explicitly connects his concept to that of Descartes for whom space, identifying itself with extension, cannot exist without bodies. In this regard Einstein said: "Descartes was not so far from the truth, when he believed he had to exclude the existence of an empty space. His conception appears absurd, as long as physical reality is seen in ponderable bodies. Only the idea of the field as a representative of reality, in combination with the general principle of relativity (which establishes the invariance of physical laws in front of every change of space-time variables, leaving invariant the square of the infinitesimal space-time interval of a photon or electromagnetic wave called $ds^2$, in a few words Pythagoras theorem), can reveal the real core of Descartes' idea: there is no empty space of interaction field.

The concept of field in a full general and absolute physical sense appears for the first time in the General Relativity, encompassing and supplanting at the same time the question of action at a distance or through a material or semi-material medium.

This concept has then be extended also to Relativistic Quantum Mechanics, but it was, and it continues to be, a concept that even today is not possible to put in axiomatic form or to derive from axioms. Rather, it can merely be described mathematically, through its action on bodies. Moreover, there is a serious problem of definition of the field that separates General Relativity, where the field exists and does not propagate, depriving the energy of a clear and unequivocal definition, from Quantum Mechanics, where the field propagates in an unambiguous way but the value of energy is not defined in an unambiguous way in all time intervals leading to temporary violations of its conservation.

FOURTH THEORY: QUANTUM MECHANICS

Quantum Mechanics finds its phenomenological foundation in the concept of duality expressed by De Broglie [29,30]: matter and electric radiation show phenomena in which they behave as waves and others in which they behave as bodies [31].

At this point it is good to give in short, the definitions of matter and electric radiation at elementary level based on their properties.

Matter: it has charge (if in particular electric can also be a globally neutral system), has mass even at zero speed, has lower speed than electric radiation.

Obviously, it is intended for matter that constituted by elementary bodies established within known limits, which for bodies with elementary electric charge, even globally null, are both the form factor (spatial extent of the distribution of elementary electric charge) and the electromagnetic radius (evaluated both statically, velocity of the body null, and dynamically, velocity of the body not null).

Electric radiation: it has no (electric) charge, has no mass at null velocity, always has the same velocity, that of electric radiation, this velocity is identified with the invariant of Special Relativity u (as long as it has the physical dimensions of the square of a velocity).

The methodological foundation of Quantum Mechanics is provided by the so-called operational definition according to Heisenberg [32], which could seem an evolution of the operational definition of physical quantities given by Thomson in the 19th century: "physical quantities are defined by the operations with which they are quantitatively measured" [33].

Obviously this operational definition of Thomson does not guarantee the physical reality of a measured quantity, that is that this quantity exists as a fundamental autonomous physical entity, examples are the temperature and the magnetic field which are auxiliary quantities that summarize fundamental phenomena: the energy of movement of elementary bodies in the matter that follows the laws of Avogadro and Boltzmann, (this for the temperature) and the electrodynamic coulombic field for the charge carriers in the matter that follows Avogadro's law (this for the magnetic field).

Operative definition of Heisenberg : Every concept of physics must be defined by a series of physical operations at least conceptually possible, it is less possible to represent atomic (and subatomic) objects by a model (even less mechanical); the representation of physical phenomena is abstract, it is done only through mathematical entities, so only observable physical quantities should be linked together, without resorting to the use of auxiliary quantities.

Starting from this last part of the operational definition, Heisenberg introduced as mathematical entities the tables of numbers of measures or matrices and had as a consequence a non-commutative mathematics, the algebra of matrices.

So far we have seen the prescriptions, dictated by De Broglie and Heisenberg, to which historically we have tried to adapt the axioms of Quantum Mechanics. But if the methodological framework is complete, the phenomenological one does not end with the concept of duality, but continues with Heisenberg's uncertainty principle, which shares with duality the property of mere phenomenological observation.

An exposition of the uncertainty principle is given here according to Caldirola et al. [34]:

it is impossible to know simultaneously, through an experimental determination, two quantities conjugated in the sense of analytical mechanics, such as position and momentum of a physical entity, with an accuracy as great as you want.

It is evident the strident contrast between the uncertainty principle, which makes explicit use of simultaneity as in Mechanics, with the Special Relativity, which has as a consequence the exclusion of simultaneity in physical phenomena, including measures, as an admitted fact valid in a general and absolute way.

This uncertainty principle has had a fundamental historical function in associating to the mathematical formalism of Quantum Mechanics, according to the operative definition of Heisenberg, a physical interpretation able to reconcile, even intuitively, the duality wave - body. On this principle is based, in the last analysis, the statistical-probabilistic interpretation of Born [35-36] of Schroedinger equation.

However in the axiomatic formulation of Quantum Mechanics is not explicitly used the uncertainty principle. In fact it can be demonstrated how the Heisenberg relations can be deduced in a general way from the properties of the operators that in Quantum Mechanics, according to the mathematical formulation by von Neumann [37], are associated with the observable quantities (subject to the definition of observability according to Dirac [38]).

At this point it arises spontaneously the conjecture if the uncertainty principle is not another form of the principle of inaccessibility according to Caratheodory [39, 40], which is nothing but an expression of the second law of thermodynamics [41, 42]: near any state of equilibrium of a system there are states that are not accessible by a reversible or irreversible adiabatic process. In the context of this conjecture the physical system observed and the observer are a single thermodynamic system, then the exchanges of energy observer-physical system occur adiabatically and in this sense fall in inaccessibility since the physical system and observer are in a state of equilibrium. This completes what has been said in the introduction of the first theory, Mechanics, in relation to natural motions that are valid for adiabatic physical systems, that is not exchanging energy with the observer.

It is amazing to realize the contradictions of this uncertainty principle; on the one hand it uses the simultaneity typical of Mechanics contradicting the Special Relativity, on the other hand it uses the adiabaticity as a property of the whole system "observer - physical system" contradicting the natural motions of Mechanics that do not exchange energy with the observer.

Finally, all this allows us to eliminate the uncertainty principle from the list of fundamental axioms; however this poses the serious problem of relationships between variables and dynamic variables that at some point we can no longer avoid.

Quantum Mechanics, as well, can be given an axiomatic foundation with trinitarian structure or three axioms plus the principle of correspondence or Ehrenfest theorem.

Here below are listed the axioms according to Bohr - Heisenberg - Born.

1° Axiom - Physical System

The states of a physical system are represented by a table of elements historically called $\Psi$ in general having one dimension and infinite elements, the $\Psi$ is a vector with infinite dimensions in a space of Hilbert vectors (legacy, only in mathematical sense, according to von Neumann of Heisenberg matrices), the state of the physical system does not change if you multiply $\Psi$ by a constant [37].

2nd Axiom - Physical quantities

Measurable physical quantities are represented by first order operators (linear), or tables of elements with 2 dimensions and infinite elements, which have the property of being equal to its complex conjugate transpose (hermitianity) and with n real eigenvalues which are the possible numerical values of the physical quantity that the operator represents, the vectors $\Psi_n$ associated with the eigenvalues are called eigenvectors and for these states of the system the quantity is well defined, that is unambiguous in the measurement of its value.

3rd Axiom - Born's measurement

The possible results of the measurement of a physical quantity each have a probability of being found, this probability is given by the squared modulus of the projection of the state of the physical system represented by the vector $\Psi$, on which the quantity is measured, with respect to the eigenstates $\Psi_n$ (eigenvectors) of the operator representing the physical quantity.

The axioms listed are the expression, as Caldirola indicates, of the probabilistic interpretation of Quantum Mechanics, it is in opposition to the stochastic one originally inspired by Langevin's ideas for nuclear phenomena. This probabilistic interpretation has been historically called of Copenhagen, or also of Bohr-Heisenberg or even "orthodox" (with abuse of definition) [43, 44].

To complete these axioms also for Quantum Mechanics, as for the Special Relativity, there is its principle of correspondence: the operators in general do not commute between them (heritage of Heisenberg matrix) and therefore do not commute the physical quantities they represent. But at the limit for the Planck constant h which tends to zero, that is for the non-discrete action, the quantities replace the operators and they can commute, so we obtain the Mechanics, where the quantities are canonically conjugate, and the limit of the principle of minimum action of Euler can also have the null value (zero) as the minimum of the action.

The 2nd Axiom seems to move away from the operational definition of Heisenberg [45], since the $\Psi$, the vector, does not represent any observable physical quantity but only an auxiliary mathematical entity, partly interpretable also as auxiliary physical quantity thanks to the 1st Axiom.

The 3rd Axiom resolves the contradiction by giving partial physical meaning of observability to a well-defined use of $\Psi$, i.e. by means of a mathematical operation: the projection which is in practice the Pythagorean theorem. We notice how Pythagoras theorem plays a fundamental physical role in the two theories, General Relativity and Quantum Mechanics.

We notice also the role of the correspondence principle in the two theories of Special Relativity and Quantum Mechanics to lead them back to Mechanics, even if using different physical objects u and h for their opposite variations and opposite limits.

It is useless to dwell in the meanders of the further internal contradictions of Quantum Mechanics, it is instead reported another axiomatic foundation according to Caldirola, which has the merit to present greater simplicity and immediacy despite the high degree of abstraction however imposed by the operational definition of Heisenberg [45].

1° Axiom - The observable

There is a complex scalar Ψ which contains in itself all the information that can be given about a particle, precisely it allows to calculate for each physical quantity, called according to Dirac observable [38], the possible values resulting from a measurement (which are obviously real) and the corresponding probability that the quantity has to assume these values at a generic instant of time.

2nd Axiom - Operator

To each observable corresponds, in the mathematical formulation, a Hermitian operator which has real eigenvalues. The set of all eigenvalues is the set of all numerical values (Heisenberg tables) that can be found in a measure of the corresponding observable. The set of eigenvectors corresponding to such eigenvalues represent the eigenstates of the system (the particle) relative to such results of a measurement process.

3rd Axiom - Evolution

The space-time evolution of a physical system is governed by the Schroedinger equation, provided that at the initial instant the state of the physical system is determined by an observation of the largest possible number of mutually compatible and independent observables, this observation is called maximum observation.

This axiomatic formulation is not free from internal contradictions, which for brevity we omit to discuss and partially resolve. However, it allows to show the principle of correspondence with Mechanics in a better form: that of Ehrenfest's theorem. Considering for a particle, intended as a physical system, its generic coordinate and its conjugate momentum, it is shown that the relations which allow to link their variations in time are obtained by the Hamiltonian operator. Since these relations are shown to be valid also for the average values of the corresponding observables. So, for the average motion of the wave packet associated with the particle (legacy of De Broglie's duality) are valid Hamilton's equations of motion and then for this motion is also valid all the Mechanics.

Obviously, all this is due to the fact that the discrete action of Planck - Sommerfeld h, which appears as a factor in the definitions of the operators, tends to zero bringing us back to the minimum value of the 3rd principle of Mechanics or the minimum action according to Euler.

It would be interesting to investigate if the principles of correspondence of Special Relativity and Quantum Mechanics are linked together in some way, certainly the axiomatic formulation of the two theories is not of great help in this investigation. In practice we could only say that Mechanics is the physical theory which, contrary to experimental evidence, is based on the approximation of an unbounded maximum causal velocity measure u, in particular as large as desired, and an unbounded minimum action measure, in particular as small as desired. Whether then to give to u the value of "infinite" implies the "infinitesimal" value of h is another matter related to the very nature of the photon and of the electric interaction.

RELATIVISTIC QUANTUM MECHANICS

Relativistic Quantum Mechanics (also conventionally called QED) is successfully applied to electric interaction and with appropriate modifications also to leptonic interaction (or weak nuclear, i.e. radioactivity). It is based on the same axioms already listed with the addition of a unique axiom derived from the 2nd Axiom of Special Relativity (but not the 2nd of General Relativity because they are not contemplated spaces other than the Minkowski flat space) that is the covariance applied to operators.

Unique Axiom: Application of Covariance.

The equations that describe a physical system are composed of operators in which the operations with respect to space and time are repeated the same number of times. That is, all the operators of the equation which allows to obtain the state $\Psi$ of the physical system have the same order and therefore are applied to the same variable entity $\Psi$ an equal number of times with respect to the same variable, whether it be space or time.

In practice this Unique Axiom is a mere completion of the 2nd Axiom in the form according to Caldirola, but of fundamental importance in determining the final success of

Quantum Mechanics and Special Relativity as theories able to fully describe the electric interaction, and the leptonic interaction as its closest relative.

At the end of this excursus in the axioms of Quantum Mechanics it is necessary to note that in any way they are considered, they violate Heisenberg's operational definition [45], even if they desperately try to adapt to it, since they are born from it. Consistency would have wanted that subject of equations, or unknowns, were only measurable quantities, in practice energy levels or frequencies of radiation emitted or absorbed by physical systems, which is the same unless Planck constant. Unfortunately, a mathematical formalism that had such variables as unknowns and reconciled the duality while being reasonably easy to use, which never hurts, has not been historically available, or even sought, unfortunately. So, the measurable quantities appear only as parameters of the solutions of the equations of Quantum Mechanics - also in its Relativistic variant -, i.e. the eigenvalues, precisely. In turn, the object of the equations remains the $\Psi$, the real unknown that, although not having the characteristic of physically existing, has the merit of being the useful mathematical tool (sic!) with which to safeguard the duality.

FUNDAMENTAL AXIOMS

After this nice race between the axiomatic foundations of physical theories, it is correct to conclude by asking which axioms can be considered useful if not indispensable in the formulation of subsequent physical theories.

As much as it can be surprising the symmetry, even if it plays a fundamental role, it cannot be taken sine cura, that is lightly, as one of the axiomatic foundations, since it rests on the experimental evidence of the conservation (as it is well put in evidence by the Noether theorem) [46,47].

Leaving aside, for the sake of brevity, the connection between the mathematical properties of space (homogeneity and isotropy) and time (homogeneity) and the principles of conservation, it must be emphasized that everything that is conserved has an experimental evidence based on upper limits, and could not be otherwise, so nothing guarantees us from future possible evidence of symmetry violations, however small in magnitude compared to the known limits.

In fact, wanting to avoid to make the axioms depend too much on the necessities of phenomenological evidence, it is appropriate to focus on some logical necessities, well aware that the axioms must also include the fact of having to deal with the experimental necessities.

The axioms will be listed and immediately after commented, it is understood that they cannot be part of an univocal system but they can be used individually or in combination with each other.

1st Axiom - Independence

The properties of space-time are independent from the properties of interactions.

This axiom is fully applied to Mechanics, to Special Relativity, to Quantum Mechanics including its evolution in the so-called standard model that describes electric and leptonic interaction, fully, and hadronic interaction but with forcing [67,68].

2nd Axiom - Solidarity

Space-time is in solidarity with interactions so that their respective properties influence each other.

This is Finzi's version of the solidarity principle as he stated in the context of General Relativity (or gravitational interaction) [69]. It is legitimate here to wonder if it can have a general value and apply to all interactions, in which case it would involve a kind of consequence or lemma: each interaction with its peculiar characteristics determines locally its own space-time structure.

3rd Axiom - Causality

Causality is understood as the postulate which states that a physical situation depends on another (univocally) and the causal research aims to discover this dependence (identification).

All this remains true also in quantum physics (mechanics), although the objects of observation, for which there is claimed to be a dependence, are different. They consist of the probability of elementary events, not of the individual events themselves.

This is the second Born formulation of causality and in its first part it applies in a formidable way to all physical theories: the univocity of causal dependence.

The second part of the 3rd Axiom is very wisely constructed to take into account the 3rd Axiom of Quantum Mechanics in the formulation given by Born himself, to give some physical meaning to the $\Psi$ necessary to duality, but also to prevent from violations of causality due to the univocal identification of the dependence, as will be commented later.

The 3rd Axiom opens the problem of univocal identification between at least two physical situations; this problem has been historically solved by phenomenology, as it was inevitable and as it was foretold, and it is stated here in the following axiom.

4th Axiom - Identification

The maximum causal velocity for all physical theories is the velocity of the electric radiation (of the photons composing it) in the absence of matter and in the presence of only the electric and gravitational fields.

This axiom is usually and historically accepted uncritically and also uncritically is automatically extended to the other two known fields that of hadronic interaction (or strong nuclear) and that of leptonic interaction (or weak nuclear). We used the expression "in absence of matter and in presence of electric and gravitational fields only" to avoid the expression "in vacuum", in addition to the well known problems that the presence of matter gives to the definition of this speed. In fact, it agrees both with the Cartesian definition given by Einstein for which the vacuum is full of field, and with the consequence of Heisenberg principles in Relativistic Quantum Mechanics for which the vacuum is full of virtual objects (renormalizable according to Feynman). Obviously, Einstein and Heisenberg conceptions of vacuum (or if you prefer Feynman), even if in contrast with each other, both clash with the need to determine the energy level of vacuum, a question atavistic since thermodynamics but that is not addressed here, having also consequences on Lorentz Invariance and conversely on physical and mathematical properties of space-time that in general could no longer be assumed homogeneous and isotropic as in Noether theorem [46,47].

But the vexed question raised by the 4th Axiom is whether the maximum causal velocity, which is identified by means of a physical entity and its particular numerical value, and which thus also respects the 3rd Axiom of Special Relativity, can disregard the 2nd Axiom, that is, the space-time structure of each interaction.

This leads us to consider three general experimental evidences to deal with in the construction of physical theories.

I Evidence

Distances in space can be measured only by exchanging energy with an interaction.

See in this regard the 3rd Axiom of Special Relativity.

II Evidence

The observer is able to exchange energy only by electric interaction.

See the uncertainty principle that is a bright example.

III Evidence

Electric interaction is the paradigm (therefore) of all phenomena due to various interactions.

This last evidence deserves in truth the title of axiom of measure, and be counted with the other four fundamental axioms.  The I and II evidence together with the fundamental axioms 2, 3 and 4 bring us back to the problem of defining which are the variable entities with which we describe the physical events and first of all the measurement of distances for the search of causality.

We then conclude by introducing the two postulates of variables.

I Postulate of Variables

There are static variables, space-time coordinates, which do not intervene in the interaction.

This obviously contradicts the 2nd Fundamental Axiom, namely the principle of solidarity.

II Postulate of variables

There are dynamic variables, space-time impulses, which intervene in the interaction (but they are measured only electrically).

In conclusion we can pose the problem if the extension of the principle of solidarity to all interactions can impose to consider all variables as dynamic; in this case we would also have the problem to reconsider the meaning of the uncertainty principle, at least in the use of Yukawa [70], in which there would no longer be a relationship between coordinates (static variables) and moments (dynamic variables), but only between dynamic variables [71] with great advantage of logical homogeneity [72].

Before closing, a comment is in order. Since our approach allows us to arrive as far as QED and even the Standard Model, it might seem surprising, at first sight, that fermions and the Pauli principle should be left out in our treatment, and an axiom like the spin-statistics connection simply ignored.

However, this choice is precisely in the spirit of the present work, which aims programmatically to avoid physical entities known only by phenomenological-experimental means and little justifiable by mathematical-theoretical means, at least until now.

For example, let us consider the fact that, in spite of the great disparity in mass and at equal apparent stability, the proton and electron, in addition to having the same amount of electric charge, do possess equal self-rotation: what is the fundamental reason for such a fact?

In fact, the connection between self-rotation, or spin, and statistics has the mere basis of ascertainment and observation, hence it cannot be put forward as an axiom, except for purposes of convenience.

Moreover, the exclusion principle and its physical foundation in the self-rotation are just based on the mere justification of the Dirac spinors, borrowed from the equation of the same name, but they are neither deducible nor axiomatizable.

For these reasons these two arguments are avoidable in the present time, that is, in the present state of knowledge.

Data availability

Data are available upon request by the corresponding Author (SB).